\documentstyle[11pt,aaspp4]{article}

\begin{document}

\input epsf

\def\cen{\centerline}
\def\nid{\noindent}
\def\vsmall{\vskip0.2in}
\def\dPhidx{{{\partial\Phi}\over{\partial x}}}
\def\dPhidz{{{\partial\Phi}\over{\partial z}}}
\def\dPhidzp{{{\partial\Phi}\over{\partial z^\prime}}}
\def\dPhidzpp{{{\partial\Phi}\over{\partial z^{\prime\prime}}}}
\def\dPhidxdx{{{\partial^2\Phi}\over{\partial x^2}}}
\def\dPhidxdy{{{\partial^2\Phi}\over{\partial x\partial y}}}
\def\dPhidydy{{{\partial^2\Phi}\over{\partial y^2}}}
\def\simgt{\hbox{\rlap{\raise 0.425ex\hbox{$>$}}\lower 0.65ex\hbox{$\sim$}}}
\def\simlt{\hbox{\rlap{\raise 0.425ex\hbox{$<$}}\lower 0.65ex\hbox{$\sim$}}}

\title{LARGE SCALE QSO-GALAXY CORRELATIONS AND WEAK LENSING}

       \author{Liliya L.R. Williams}
       \affil{Department of Physics and Astronomy\\  
              University of Victoria\\
              Victoria, BC, V8P 1A1, Canada}
       \authoremail{llrw@uvastro.phys.uvic.ca}

\begin{abstract}
Several recent studies show that bright, intermediate and high redshift 
optically and radio selected QSOs are positively correlated with nearby 
galaxies on a range of angular scales up to a degree. Obscuration by 
unevenly distributed Galactic dust can be ruled out as the cause, leaving
weak statistical lensing as the physical process responsible. However the 
amplitude of correlations on $\simlt 1^\circ$ angular scales is at least 
a factor of a few larger than lensing model predictions. A possible way to 
reconcile the observations and theory is to revise the weak lensing formalism.
We extend the standard lensing formulation to include the next higher order 
term (second order) in the geodesic equation of motion for photons. We derive 
relevant equations applicable in the weak lensing regime, and discuss 
qualitative properties of the updated formulation. We then perform 
numerical integrations of the revised equation and study the effect of the 
extra term using two different types of cosmic mass density fluctuations. 
We find that {\it nearby large-scale coherent} structures increase the 
amplitude of the predicted lensing-induced correlations between QSOs and 
foreground galaxies by $\sim 10\%$ (not a factor of several required by 
observations), while the redshift of the optimal, i.e. `most correlated'
structures is moved closer to the observer compared to what is predicted 
using the standard lensing equation.
\end{abstract}

\section{Introduction}

Weak gravitational lensing manifests itself in two observable ways: if a 
source is extended, like a galaxy, its image will appear to be sheared 
tangentially with respect to the foreground mass concentration. If sources 
are unresolved, like QSOs, weak lensing can be detected in a statistical 
way by the sources' angular (anti)correlation with the tracers of the 
intervening mass distribution. 

Statistical QSO-galaxy associations on large angular scales, $10'-1^\circ$ 
are the subject of the present paper. The mechanism responsible for the 
associations is believed to be the magnification bias, which depends on the
shape of the number-magnitude counts of the background sources. If the 
number counts are a power law with a slope $\alpha=d$log$N(m)/dm$ then 
the over- or underdensity of sources down to a certain limiting flux 
behind the lens is related to the lens magnification $M$ by 
$q(M,\alpha)=M^{2.5\alpha-1}$. The factor $2.5\alpha$ in the exponent
accounts for the magnification of individual sources brightened into
a flux limited sample, while the factor $M^{-1}$ corrects for the area
dilution of the number density of sources on the sky. Thus if $\alpha>0.4$ 
($\alpha<0.4$) correlations (anticorrelations) are predicted between 
galaxies and background sources.

There are a number of observations of correlations between various
types of QSOs and foreground galaxies. Here is a sampling of the recent 
literature. Radio selected 1 Jy QSOs at $z_s>0.5$ (K\"{u}hr et al. 1981) 
are correlated with the APM Catalog galaxies, $z_l\sim 0.1-0.3$,
(Seitz \& Schneider 1995, 
Ben\'{\i}tez \& Mart\'{\i}nez-Gonz\'alez 1995, 1997, Norman \& Williams 
1999) on scales between a few arcminutes and a degree. Bright optically 
selected QSO from the LBQS Catalog (Hewett et al. 1995) are correlated with 
the APM galaxies on degree scales (Williams \& Irwin 1998), while faint 
UVX selected QSO candidates are anti-correlated with galaxy groups on 
$\simlt 10'$ scales (Croom \& Shanks 1999). The amplitude of 
(anti)correlations varies between $30\%$ and $1-2\%$ depending on the 
angular scale and the details of each study.

As judged by their redshifts, QSOs and galaxies in all these studies are
unconnected, therefore there are two possible physical reasons why these 
would appear to be associated in projection. One is weak lensing, as we have 
just described, the other is dust obscuration.  However, a number of recent 
studies, if considered together, reject dust as the likely explanation. 

Depending on the limiting magnitude of the QSO sample magnification bias
will induce foreground galaxies to be either positively or negatively 
correlated with QSO. Bright end of the QSO number counts has a steep slope,
$\alpha>0.4$, and so $q>1$; the opposite is true for QSO samples with faint
limiting magnitudes, where  $\alpha<0.4$.
And in fact $q$ is observed to be $>1$ in most of the above listed
studies because most of them use bright QSOs. As one goes to
fainter QSOs $q$ seems to drop to $\sim 1$, while faint QSOs, those in
the Croom \& Shanks study are anticorrelated with the 
intervening mass, i.e. $q<1$. However to explain $q>1$ and $q<1$ 
cases with dust obscuration one needs to invoke two different types of
dust: Galactic dust will obscure both QSOs and galaxies in some directions
thus leading to positive correlations between the two populations, while
dust intrinsic to lenses, i.e. galaxy clusters/groups will obscure QSOs
in the direction of the lenses only, leading to anticorrelations. This
double-type effect makes dust a rather contrived scenario.  

Furthermore, if Galactic dust is responsible for the observed positive
correlations one would expect the radio selected QSO samples to be
less affected, and thus show smaller cross-correlation signal compared
to optically selected QSOs. The observational situation is exactly opposite: 
1 Jy QSOs are stronger correlated with APM galaxies (Norman \& Williams 1999) 
than are LBQS QSOs (Williams \& Irwin 1998) on similar angular scales. 

Qualitative evidence not only rejects dust as the underlying physical
cause, but also points to lensing. Amplitude and significance
of correlations generally increase with brighter QSOs, as predicted by
magnification bias. QSOs at intermediate redshifts, $z_s\sim 1-2$, i.e.
those roughly at the optimal distance for being lensed by structures at
$z_l\sim 0.1-0.3$ show the strongest correlations. Dust obscuration
should show no preference for any particular QSO redshift range.

All evidence combined strongly suggests that lensing is responsible for 
the observations. In fact, lensing would have been long accepted as the 
cause were it not for the unexpectedly large amplitude of the
(anti)correlations. On small angular scales, $\simlt 10'$ the non-linear
growth of mass fluctuations with cosmic time produces the correct correlation
strength; Dolag \& Bartelmann (1997) and Sanz et al. (1997) reproduce the 
1 Jy--APM correlations observed by Ben\'{\i}tez \& Mart\'{\i}nez-Gonz\'alez 
(1995, 1997). However, for correlations on larger scales, $\simlt 1^\circ$, 
and anticorrelations on $\simlt 10'$ scales (Croom \& Shanks 1999) the 
strength of associations exceeds predictions by about a factor of 5-10.

There are three avenues within the lensing hypothesis for reconciling 
the model predicted and observed correlation amplitudes. One is to make 
the slope of the QSO number counts in the appropriate redshift range
very steep. Williams \& Irwin (1998) estimate the required slope $\alpha$ 
to be about 8. Because weak lensing induces small magnifications, 
the number counts slope of `lensed' QSO, i.e. those seen in the directions
of intervening mass concentrations should be almost as steep as the slope of
the `unlensed' QSOs. In other words, the overall QSO number counts slope 
should also be around 8. This is grossly inconsistent with the observed slope 
of $\sim1.1-1.6$. We consider this option the least likely of the three. 

The second alternative is to invoke mass density fluctuations on 
$\simlt 2-10h^{-1}$Mpc scales ($10'-1^\circ$ at $z_l\sim 0.1-0.3$) that are 
substantially larger than observations seem to indicate. Williams \& Irwin 
(1998) and Croom \& Shanks (1999) independently estimate that 
$\Omega\sigma_8$ should be about 3-4 to explain their respective results. 
The commonly accepted value for $\Omega^{0.6}\sigma_8$ is $\approx 0.6$, and
is derived by two different methods: bulk flows in the nearby Universe and 
the abundance of rich galaxy clusters (Branchini et al. 1999, 
White et al. 1993). While scale-dependent biasing may allow one to increase 
the amplitude of mass fluctuations by $10-50\%$, an increase by a factor of 
6-8 is outside the range of acceptable possibilities. 

The third option is that the weak magnification regime of the lensing theory 
needs a revision. If future surveys support the results of present 
observations then this option will have to be considered seriously. 

In principle many aspects of the currently accepted description of light
propagation in an inhomogeneous medium can be questioned, for example the 
applicability of the Friedmann-Robertson-Walker metric and its weakly 
perturbed version in cosmological context, 
the validity of the general relativistic geodesic equation of motion as 
applied to light rays, etc. But the overwhelming success of the standard 
picture, which incorporates these concepts, in describing our Universe in 
general and observations of most types of gravitational lensing in 
particular suggests that the modifications to the standard description 
of lensing has to be sought within the existing framework.
The standard lensing equation is the solution to the first order 
approximation of the full geodesic equation of motion for photons. There 
have been attempts in the literature to use the full geodesic equation to 
propagate light rays through a clumpy universe (Holz \& Wald 1999, Tomita 
et al. 1999 and references therein), however these studies were focused on
small angular scale lensing effects, and the equation was used in conjunction
with a limited range of prescriptions for mass fluctuations, and thus not
fully explored.

Instead of taking on the full geodesic equation with all the higher order 
terms, our aim in this paper is to go one step beyond the standard lensing 
description, i.e. study the effect of the second order term(s) in the 
geodesic equation. The advantage of this approach is that we can derive 
useful analytic results, for example, an approximate form of the revised 
lensing equation (Section 3), and the equation for the increment 
in source magnification arising from retaining the second order term(s) 
(Section 4), and discuss qualitative predictions of the revised formulation
(Section 5). With the help of numerical integrations we examine how the 
effects of the second order term(s) depend on the type of mass fluctuations 
populating the universe (Section 6 and 7).

\section{Light propagation equation with second order terms included}

The derivation of the standard lensing equation can be found in many
places in the literature (Schneider et al 1992, and references therein;
Dolag \& Bartelmann 1997); here we follow the derivation presented by
Kaiser (1998), but for simplicity adopt a spatially flat cosmological model.
Unlike Kaiser and other authors we do not truncate the geodesic equation
at the first order terms, but retain the second order term(s) as well.
 
The FRW spacetime metric with superimposed weak perturbations is given by
$$ds^2=g_{\alpha\beta}dr^{\alpha}dr^{\beta}=
a^2(\eta)[-(1+2\Phi)d\eta^2+(1-2\Phi)(dx^2+dy^2+dz^2)],\eqno(1)$$
where $a(\eta)$ is the scale factor, $\Phi$ is the Newtonian potential, 
$\eta$ is the conformal time related to cosmic time, $t$ though
$dt=ad\eta$, and $dx$, $dy$, $dz$ are the comoving separations between
two adjacent points in space.
Photon trajectories are solutions of the geodesic equation,
$${{d^2r^\alpha}\over{d\lambda^2}}=-g^{\alpha\beta}
\Bigr(g_{\beta\nu,\mu}-{1\over 2}g_{\nu\mu,\beta}\Bigl)
{{dr^\nu}\over{d\lambda}}{{dr^\mu}\over{d\lambda}},\eqno(2)$$
where $\lambda$ is the affine parameter, and $r^{0,1,2,3}$ stand for
$\eta, x, y$, and $z$ (redshift is designated by $z$ with a subscript,
either $z_l$ or $z_s$). The elements of the diagonal metric tensor
$g_{\alpha\beta}$ can be read off from the line element, eq.(1). Let us 
write down the $x$-component of the  geodesic equation keeping terms up to 
second order in small quantities, i.e. $\Phi$,
$dx/d\lambda$, and $dy/d\lambda$:
$${{d^2x}\over{d\lambda^2}}\approx
-{{\partial\Phi}\over{\partial x}}\Bigl({{d\eta}\over{d\lambda}}\Bigr)^2
-{{\partial\Phi}\over{\partial x}}\Bigl({{dz}\over{d\lambda}}\Bigr)^2
+2{{\partial\Phi}\over{\partial z}}
\Bigl({{dx}\over{d\lambda}}\Bigr)\Bigl({{dz}\over{d\lambda}}\Bigr).
\eqno(3)$$
The first two terms in this equation are the standard first order terms; 
the last term is the only surviving second order term. 
Changing the differentiation throughout to that with respect to $z$ and
using that $d\eta/dz=1-2\Phi$ to first order, we get
$${\ddot x}\approx-2\dPhidx+2\dPhidz{\dot x},\eqno(4)$$
where the dots represent differentiation with respect to $z$, and
$x$ is the transverse comoving separation between a ray and the $z$-axis. 
A similar expression can be written for the $y$-component. This is the 
equation of motion for the $x$-component of a single light ray as it
wonders through a spatially flat universe populated with small mass
perturbations. The standard equation (see eq.[8] of Kaiser 1998) is eq.(4) 
minus the last term.

An alternative way to arrive at eq.(4) is by starting from Fermat's
Principle which states that the path taken between two points in space, 
$P_1$ and $P_2$, is a local extremum, 
$$\delta\int_{P_1}^{P_2}d\eta=0.\eqno(5)$$
From the metric of eq.(1) and the condition that light rays are null 
geodesics we get
$$d\eta=(1-2\Phi)(dx^2+dy^2+dz^2)^{1/2},\eqno(6)$$
which can be rewritten as
$$n|{\bf{t}}|=1,\eqno(7)$$
where $n=(1-2\Phi)$ is the refractive index of the space-time described
in eq.(1),
${\bf{t}}=\dot{x}\hat{\bf x}+\dot{y}\hat{\bf y}+\dot{z}\hat{\bf z}$ is
the unit velocity vector of the light ray, and the dot represents $d/d\eta$. 
Equation (5) can now be rewritten as
$$\delta\int_{P_1}^{P_2}n|{\bf{t}}| d\eta=0.\eqno(8)$$
The integrand is an extremum when Euler-Lagrange equations are
satisfied, so in the $x$-direction we must have
$${d\over{d\eta}}\Bigl[{{\partial(n|\bf{t}|)}\over{\partial\dot{x}}}\Bigr]
-{{\partial(n|{\bf{t}}|)}\over{\partial x}}=0,\eqno(9)$$
which is simplified to a differential equation for light rays:
$${d\over{d\eta}}\Bigl[n{\bf{t}}\Bigr]-\nabla n=0.\eqno(10)$$
Rewritting this as
$$n{\dot {\bf t}}=\nabla n-{\bf{t}}{d\over{d\eta}}n,\eqno(11)$$
makes it apparent that we have arrived at the vector form of eq.(4).
(This is the same as eq.[4.17] of Schneider et al. [1992], without the 
last term in their equation.)
The term $\nabla n$ is the gradient of the refractive index
evaluated in the coordinate system whose $z-$axis is
aligned with the unperturbed light ray, while the term
${\bf{t}}\,{d n/d\eta}$ is along the direction of the actual ray.
Thus the RHS of eq.(11) is the gradient of $n$ in the plane
perpendicular to the direction of the actual ray.

Because the last term in eq.(11) (i.e. the second order term) is small 
compared to the first one, 
is it neglected in the standard lensing equation. In physical terms
this amounts to setting $\dot x$ and $\dot y$ to zero in eq.(11), so
that $\bf{t}$ is along the $z-$axis. In other words, this approximation, 
sometimes called the Born approximation, assumes that the actual ray 
is always parallel to the unperturbed ray. In this paper we use the full
eq.(11) and thus test the validity of the Born approximation.

Using (4) we now write down the comoving separation between two adjacent 
rays, say rays on the opposite sides of a small light bundle:
$$\Delta{\ddot x}\approx -2\Delta x\dPhidxdx-2\Delta y\dPhidxdy
                         +2\dPhidz\Delta {\dot x},\eqno(12)$$
and similarly for the $y$-component. Here, the terms
$2\Delta x{{\partial^2\Phi}\over{\partial z\partial x}}\dot x$ and
$2\Delta y{{\partial^2\Phi}\over{\partial z\partial y}}\dot x$
are much smaller than the third term on the right hand side in the above 
expression, so they were omitted.

\section{Approximate solution: revised lensing equation}

Equation (4) without the second order term has an exact solution,
$$x=\theta_0 z-2\int_0^z\dPhidx (z-z^\prime)dz^\prime,\eqno(13)$$
where $\theta_0$ is the observed angle of the source with respect to
the $z$-axis in the absence of perturbations. If all the mass 
fluctuations between the observer and the source are confined to a plane 
perpendicular to the $z$-axis then (13) reduces to the standard lensing 
equation. There is no useful solution for the full equation (4). 
To find an approximate solution in the weak lensing limit let us assume 
that $\dot x$ is constant:
$$\dot x\approx x/z\approx\theta_0.\eqno(14)$$
This is not a bad approximation in the weak lensing case where path
deviations suffered by light rays are small. With this assumption the 
second order term in eq.(4) can be integrated by parts,
$$2\dot x
\int_0^zdz^\prime\Bigl( \int_0^{z^\prime}dz^{\prime\prime}\dPhidzpp\Bigr)
 =2\dot x\Biggl(
z\int_0^z\dPhidzp dz^\prime-\int_0^z \dPhidzp z^\prime dz^\prime\Biggr)
 =2\dot x \int_0^z\dPhidzp(z-z^\prime)dz^\prime.\eqno(15)$$
Thus an approximate solution to equation (4) is
$$x\approx \theta_0 z-2\int_0^z\dPhidx (z-z^\prime)dz^\prime
           +2\dot x \int_0^z\dPhidzp(z-z^\prime)dz^\prime.\eqno(16)$$
This is the revised version of the `3-dimensional' lensing equation (13).
The equation for the transverse separation between two adjacent rays
is then,
$$\Delta x\approx~\Delta\theta_0 z\Bigl[1
          -2\int_0^z\Bigl(\dPhidxdx+\dPhidxdy\Bigr)
                   ~{{z^\prime(z-z^\prime)}\over z}~dz^\prime
          +2\int_0^z\dPhidzp  ~{{(z-z^\prime)}\over z}~dz^\prime\Bigr],
\eqno(17)$$
where we have approximated $\Delta\dot x$ by $\Delta\theta_0$, and set
$\Delta x\approx\Delta y\approx\Delta\theta_0 z^\prime$, similar to eq.(14),
i.e. angular separation between two adjacent rays is constant to zeroth 
order. 

\section{ Magnification of sources}

If the source does not suffer much distortion and the $x$ and $y$-axes 
are along the major and minor axes of the image, then the magnification of
the image relative to the source is approximately 
$$M\approx \Bigl({{z\Delta\theta_0}\over{\Delta x}}\Bigr)
           \Bigl({{z\Delta\theta_0}\over{\Delta y}}\Bigr).\eqno(18)$$ 
The standard magnification in the weak lensing regime is then given by
$$M_{std}\approx 
         1+2\int_0^z\Bigl(\dPhidxdx+\dPhidydy\Bigr)
         ~z^\prime~(1-{{z^\prime}\over z})~dz^\prime
\eqno(19)$$
and the difference between $M_{std}$ and magnification with the second
order term is,
$$\Delta M=M_{2nd}-M_{std}\approx
-4\int_0^z\dPhidzp~(1-{{z^\prime}\over z})~dz^\prime.\eqno(20)$$
This additional magnification depends on the gradient of the Newtonian 
potential along the $z$-axis, or along the line of sight to the source.
Equation (20) was found to be in excellent agreement with 
the results of numerical integrations (see below) for magnifications 
$M_{std}\simlt 1.3$. This range is more than sufficient for the present 
study; magnifications typically produced by smoothed large scale 
structures on $\simlt 1^\circ$ scales are less than $1\%$. For example,
mass fluctuations associated with the $18.5\leq m_R\leq20.0$ APM galaxies
in Williams \& Irwin study are of the order 1.005.

\section{Properties of the equation}

Qualitative consequences of the second order term in the geodesic equation
can be assessed from eq.(17) and (20). 
Let us consider a single positive mass fluctuation between us and 
the source. $\partial\Phi/\partial z$ will be negative between us
and the lens, and because of the $(1-z^\prime/z)$ weighting in the last
term of eq.(17) it will decrease $\Delta x$, i.e. increase magnification. 
The opposite is true for fluctuations of negative $\delta\rho/\rho$. 
Alternatively, looking at eq.(11) we see that the second order term 
increases the bending of the rays around positive and negative potential 
wells beyond what is predicted by the standard first order term.
Thus the second order term reinforces the amplitude of (de)magnification 
predicted by the standard formulation. 

According to eq.(17) both $\Delta x$ and $\Delta y$-components of the beam's 
cross section are affected in the same way by the second order term,
i.e. a factor proportional to 
$\int_0^z\dPhidzp~(1-{{z^\prime}\over z})~dz^\prime$
is added to both, therefore magnification along the 
$x$-direction is not at the expense of magnification in the $y$ direction,
and so for small magnifications there will be no change in the 
lensing-induced shape distortion of the image. This was verified by the
numerical integrations of eq.(12).

When a significant lens is encountered, meaning the projected mass 
density fluctuation is high, the middle term in eq.(17) becomes large, as it 
is related to the mass density through the Poisson equation. Thus the 
magnification is dominated by the second derivatives of $\Phi$, and the 
extra magnification supplied by the second order term, $\Delta M$ is
negligible compared to $(M_{std}-1)$. In other words the standard 
equation is completely adequate in the regions of high density, for 
example in galaxy clusters and near individual galaxies.

An important aspect of the standard lensing equation is that not all
equal amplitude mass fluctuations between the observer and the source 
contribute equally to the source magnification; eq.(19) shows that the 
fluctuations are weighted by $z^\prime(1-z^\prime/z)$, i.e. lenses 
approximately half way between the source and the observer are most 
influential. $\partial\Phi/\partial z$ in eq.(17) and (20) is weighted by 
$(1-z^\prime/z)$, thus the lenses closer to
the observer contribute more through $\Delta M$. In practice it is
probably the combination of the amplitude of fluctuations, their physical
scale and proximity to the observer that determines their relative
contributions to $\Delta M$. For example, because $(1-z^\prime/z)$ 
depends on the distance only weakly high frequency fluctuations will 
mutually cancel out, and will not contribute significantly to $\Delta M$.
$\Delta M$ is probably dominated by the `last screen' of large scale
structure, like superclusters, voids and filaments, located closer to the
observer than the standard optimal lensing distance.

This is an interesting point in the light of some of the observational
results.
A few cross-correlation studies seem to indicate that the galaxies most
strongly correlated with the QSOs are not the ones at the optimal lensing
redshifts but are somewhat closer than that to the observer. This was 
stressed by Bartelmann \& Schneider (1993) who reanalyzed Fugmann (1990)
result of correlations between Lick Catalog galaxies and 1 Jy sources.
It was also found to be the case by Williams \& Irwin (1998) who show
that correlations are strongest for $z_{QSO}\geq 1$. The optimal lens
location in the Einstein-de Sitter universe model for sources at 
$z_{QSO}\geq 1$ would be 0.4, while the galaxies in their study lie at 
$z_l\sim 0.2$. 

Equation (20) for $\Delta M$ shows that flux is conserved with the 
addition of the second order term, as $\partial\Phi/\partial z$ will 
average out to zero for a collection of sources. 

Finally, because the second order term depends on $\partial\Phi/\partial z$ 
it is essential that the continuous, 3-dimensional nature of the mass 
fluctuations is taken into account. If the effect of the second order term 
proves to be non-negligible then the treatment of weak QSO lensing based on 
multiple-plane formalism, where the standard lensing equation is solved 
successively on many 2D planes stacked perpendicular to the optical axis, 
should not be used.

We will return to some of the qualitative remarks made in this Section in 
Section 7, where we discuss the results of numerical integrations of eq.(12).

\section{Application of the equation}

\subsection{Numerical integrations}

To study the effect of the second order term further we need to consider
an inhomogeneous mass distribution in the Universe, solve eq.(12) for many 
lines of sight and compute relevant output quantities, like distribution of 
magnifications, average magnification, etc. Given the simplicity of the 
approximate solution, eq.(17), this can be done analytically for simple mass 
distributions. For an arbitrary mass distribution a numerical integration 
approach is preferred. We use 5th order Runge-Kutta method to integrate 
eq.(12) and its $y$-component counterpart. In principle, as a light bundle 
propagates through a clumpy medium the path of its central ray will deviate 
from the fiducial ray in the absence of perturbations; however, one can 
ignore this effect in the weak lensing regime and evaluate the gradients of 
the Newtonian potential in eq.(12) along the fiducial central ray. 
When integrating eq.(12) we use boundary conditions at the observer;
for every line of sight we start with two images seen by the observer, each 
one confined to the $x$- and $y$-axis entirely, and propagate these backwards 
to the redshift of the source. The shape, size and orientation of the two 
final sources per line of sight allows us to solve for the four components 
of the total magnification matrix. All sources are located at $z_s=1.5$, 
typical redshift of QSOs in correlation studies.

\subsection{Mass density fluctuations}

We are primarily interested in how the weak lensing model predictions are 
affected by the inclusion of the second order term,  in particular if the
differences in the standard vs. revised predictions depend on the type of
mass density fluctuations present in the universe. Therefore we pick two 
rather different, but simple mass fluctuation scenarios: randomly distributed 
spherically symmetric mass clumps, and a Gaussian random-phase field. 

In both the scenarios we limit ourselves to large scale structures only, and 
do not attempt to include the lensing effects of smaller, more compact 
objects like individual galaxies and galaxy clusters. The computed 
correlation amplitude will not be affected by this exclusion: 
Theoretical calculations of weak lensing induced correlations effectively 
smooth out density fluctuations on scales below the correlation scale.
This is seen clearly from equation (13) of Dolag \& Bartelmann (1997),
where the cutoff is imposed by the filter function $F(k,\phi)$. 

On large scales the linear growth of mass density fluctuations is an 
adequate assumption; therefore at every cosmic epoch we scale the amplitude 
of fluctuations by $a(t)=1+z_l$.

\subsubsection{Model \#1: Clumps}

This model for the mass distribution is motivated by the observation that 
the large scale structure looks like a network of walls, voids and
filaments, i.e. coherent structures spanning tens of Megaparsecs. 
Therefore we represent the mass fluctuations by a collection of
many spherically symmetric clumps. We have experimented with a number of 
different density profiles, including non-singular isothermal sphere, 
top hat, and Hernquist profile, $\rho\propto [r(1+r/r_0)^3]^{-1}$.
Qualitatively the results described below are similar for all types of mass 
clump profiles provided the physical scale of the clumps is about the same.
In what follows we use clumps described by a Newtonian potential,
$\Phi(r) \propto -e^{-r/r_c}$, where $r$ is comoving distance from the
clump center, and $r_c$ is its scale length.
This particular shape was chosen because the potential falls off quickly 
so that the influence of the mass clumps does not extend over scales 
comparable to the Hubble length. The mass profile of a clump is 
$$\rho(r) \propto {{e^{-r/r_c}}\over{r r_c}}~\Bigl(2-{r\over r_c}\Bigr).
\eqno(21)$$
The profile has a central mass peak ($r<2r_c$) and is surrounded by a 
shallow region of the opposite density. The clumps are quite compact: 
for $r_c=0.1 {c\over{H_0}}$ the density drops by a factor of 
$\sim 50$ between $r_c/10$ and $r_c$, and by a factor of $\sim 20$ 
between $r_c$ and $3r_c$. The mass enclosed within $r$ of each clump goes
as $e^{-r/r_c}r^2 r_c^{-1}$, and asymptotically approaches zero. We use
clumps of positive and negative central mass density, so even in a small 
volume the statistical expectation value for the net excess mass density 
is zero. This is in accord with the equations derived in Sections 2, 3 and 
4 which assume a flat space geometry. 

The clumps are randomly distributed in space with a typical interclump 
separation of about $70h^{-1}$Mpc. The line of sight from the observer to 
the source is completely immersed in a sea of clumps. All clumps are 
identical, with $r_c=0.1 {c\over{H_0}}=300h^{-1}$Mpc which is larger than the 
interclump separation. Because of this the typical radius of the net 
structures formed by the superposition of clumps is about 
$0.05 {c\over{H_0}}=150h^{-1}Mpc$, comparable to the size of the largest
superclusters seen in the nearby Universe (Batuski et al. 1999).
The rms fluctuations of mass in spheres of $R=300h^{-1}$Mpc is 0.012, 
comparable to, or somewhat higher than the standard $\sigma_8=1$ normalized
CDM value. The amplitude of fluctuations in this model grows slowly with 
decreasing scale; the fractional contribution to $\sigma_8$ is small. We will 
return to the discussion of the amplitude of mass fluctuations later, in 
Section 7. In the redshift range $z_l=0.1-0.3$, the location of the galaxies 
used in the cross-correlation studies, the average projected separation 
between two adjacent clumps is $25h^{-1}$Mpc, or about $2.5^\circ$, comparable
to the angular scale of observed QSO-galaxy cross-correlations.

\subsubsection{Model \#2: Gaussian random field}

Here we assume that the fluctuations are represented by a Gaussian random 
field. The form of the power spectrum was taken form Peacock (1997). 
To make a fair comparison with the clumps model we limit the power spectrum 
to a range of wavenumbers between $k_{low}=2\pi(2 r_c)^{-1}=0.01h$Mpc$^{-1}$ 
and $k_{high}=2\pi(0.5 r_c)^{-1}=0.04h$Mpc$^{-1}$. In this region
the shape of the power spectrum is $P(k)\propto k$. The phases of the 
various modes are uncorrelated. 

\section{Results of numerical integrations}

\subsection{General considerations}

For every line of sight numerical integration of eq.(12) yields source
magnification. Since we are interested in the cross-correlation between 
sources and the lensing mass, we first need to relate the source 
magnifications to the cross-correlation function. 

Williams and Irwin (1998) derived a relation, applicable in the weak lensing 
regime, between QSO-galaxy correlation function, 
$\omega_{QG}(\theta)$, and the autocorrelation function of the foreground
galaxies used in the analysis, $\omega_{GG}(\theta)$ (their eq.[6]):
$\omega_{QG}(\theta)\approx(2\tau/b)(2.5\alpha-1)\omega_{GG}(\theta)$,
where $\tau$ is the total optical depth of the lensing slab of matter traced
by the APM galaxies, and $b$ is the bias parameter. The factor $(2\tau/b)$ is 
the proportionality coefficient between $(M-1)$, `magnification excess' of a
source, and normalized galaxy number excess, $(\sigma-1)$, both $M$ and 
$\sigma$ referring to the same patch of the sky. In the weak lensing regime 
$(M-1)$ and $(\sigma-1)$ are linearly related in both the standard and 
revised lensing formulations, but the coefficient is different. Thus for 
either formulation we can write 
$$\omega_{QG}(\theta)\approx
   \langle(M-1)/(\sigma-1)\rangle(2.5\alpha-1)\omega_{GG}(\theta),\eqno(22)$$
where the brackets indicate the typical value over many lines of sight. 
Without evaluating $\omega_{QG}(\theta)$ itself, eq.(22) allows us to derive 
an expression
for the change in the predicted correlation function resulting from adding 
the second order term to eq.(4), while keeping the galaxy distribution, 
$p(\sigma|\theta)$ and the slope of source number counts, $\alpha$ the same:
$${{\omega_{QG,2nd}(\theta)}\over{\omega_{QG,std}}(\theta)}
     =\Bigl\langle{{(M_{2nd}-1)}\over{(M_{std}-1)}}\Bigr\rangle.\eqno(23)$$

The quantity $(M_{2nd}-1)/(M_{std}-1)$ for any line of sight has an 
interesting property of being independent of the {\it amplitude} of mass 
density fluctuations, while being sensitive to the {\it shape} of the 
fluctuations. This is readily seen from eq.(19) and (20): both versions of 
$(M-1)$ are proportional to $\Phi$. Therefore the change in $\omega_{QG}$ is 
the same regardless of the actual amplitude of fluctuations or the 
magnifications induced by the mass structures, so long as their shape
remains the same. 

Although the actual values of magnifications are not important, the
values presented below are comparable to what one expects from
the smoothed large scale structure, i.e. $\simlt 1\%$. 

All the integrations described below conserve flux. Flux is considered
to be conserved if the ratio of the
total amount of flux received by all observers distributed in a spherical
shell around a source to the total flux emitted by that source is 1. For 
any given observer that ratio is just $1/M$; the average $1/M$ over a 1000
or so observers in each set of integrations was found to be 
less than one standard deviation of the mean away from unity.

We now describe the results for each of the two mass fluctuation models.

\subsection{Mass density fluctuations}

\subsubsection{Model \#1: Clumps}

Each point in the top panel of Figure 1 represents the magnification of 
a source computed using the standard and second order formulations: 
$M_{std}$ and $M_{2nd}$, respectively. It is apparent that including
the second order term has a small but non-negligible effect on the
computed magnifications. The bottom panel, which shows the residuals
$\Delta M=M_{2nd}-M_{std}$ vs. $M_{std}$, demonstrates that $M_{2nd}$ is,
on the average, more extreme than $M_{std}$, in other words, the second 
order term increases the computed magnification of the source if 
$M_{std}>1$, and decreases it if $M_{std}<1$. This effect was already noted
in Section 5. The fractional change is 
not very large; on the average $\Delta M/(M_{std}-1)\approx 0.2$.

Do these results alter the predicted amplitude of QSO-galaxy correlation 
function? Figure 2 shows $(M_{2nd}-1)/(M_{std}-1)$ as a function 
of the projected angular distance between the source and the nearest mass 
clump of either positive or negative central density. The only clumps 
considered here are in the redshift range $0.1-0.3$, to simulate the observed 
situation. The angular scale of correlations includes the observational
scale of $\simlt 1^\circ$.  The median of all the $(M_{2nd}-1)/(M_{std}-1)$
points is 1.1,  i.e. the amplitude of the cross-correlation function is 
increased by $10\%$ compared to the standard lensing equation. As the 
reader will remember from the Introduction, an increase by a factor of
$\simgt 5$ is needed to reconcile observations with theory. 

The solid points in the plot represent sources (lines of sight) with 
$M_{2nd}>1.001$, an arbitrary cutoff designed to separate out about half the
magnification values. The median of the solid points is very similar to the 
one quoted above, i.e. 1.1, but the dispersion in the points is much
smaller, and so it is more obvious that the points prefer to lie above
the $(M_{2nd}-1)/(M_{std}-1)=1$ line. In a simple case where the source 
number counts truncate at a flux just below our detection limit the solid 
points will represent a flux limited sample of observed sources.

Note that about half of all the points in this plot represent $M_{std}<1$, 
i.e. demagnifications. In such cases $\Delta M$ values tend to be less than
zero, therefore $(M_{2nd}-1)/(M_{std}-1)>1$, and lensing induced 
anti-correlations are also enhanced by the second order term. 

As was pointed out in Section 5 with the help of eq.(20), the presence of 
the second order term in eq.(4) may change the redshift of the optimal 
lenses compared to the
standard formulation; it may `move' the most effective lenses, for fixed 
source redshift, closer to the observer. The numerical integrations 
indicate such an effect. We have divided the nearby universe into four
slices oriented perpendicular to the line of sight to the sources.
The slices are: $z_l=0.05\rightarrow 0.1$, $0.1\rightarrow 0.2$, 
$0.2\rightarrow 0.3$, and $0.3\rightarrow 0.4$. In each slice we 
compute the median value of $(M_{2nd}-1)/(M_{std}-1)$ for sources found 
within angle $\theta$ of at least one clump. Only clumps of positive central
density were used in this exercise to emulate the observations where QSOs 
are (anti)correlated with regions of galaxy excess. The results are plotted 
in Figure 3. Angle $\theta$ was set to $0.5^\circ$ (solid), $0.75^\circ$ 
(dotted), and $1^\circ$ (dashed), comparable to the angular scale of 
observed correlations. Mass concentrations in all redshift slices are 
more correlated with the background sources than the standard lensing
formalism would predict, however, the increase in the 
correlation amplitude is larger for $z_l\leq 0.2$ slices than for the 
$z_l\geq 0.2$ slices, i.e. the inclusion of the second order term moves the 
redshift of the optimal (most correlated) mass fluctuations closer to
the observer.

How do the results change if we adopt a different value for the clump 
length scale, $r_c$? 
The effect of the second order term, evaluated with
$(M_{2nd}-1)/(M_{std}-1)$ declines rapidly  as $r_c$ is reduced by a 
factor of 3 or larger. This is easy to understand; the effects of small 
scale density fluctuations on the net $\partial\Phi/\partial z$ cancel out 
as eq.(20) is integrated between the source and the observer. Only the
effects of larger scale fluctuations, however small in amplitude, will
remain. 

\subsubsection{Model \#2: Gaussian random field}

Figure 4 compares $M_{std}$ with $M_{2nd}$ for the Gaussian random field
mass fluctuation prescription, and is equivalent to Figure 1 of the clumps
model. Though Gaussian random field does respond to the second order term, 
the change in magnification is very much smaller than in the clumps scenario,
therefore there will be virtually no increase, over the standard lensing
formulation, in the predicted amplitude of cross-correlations if a Gaussian 
random-phase field describes the mass density fluctuations on large length 
scales. 

Why does the inclusion of the second order term make a difference in the
weak lensing model predictions when a clumps density model is used, while no 
effect is seen when the density field is Gaussian? 
As we have already mentioned the normalization of the power spectrum has
no effect on $(\omega_{QG,2nd}/\omega_{QG,std})$. If normalization is
increased the magnification labels on both the axis of Figure 4a will
change, but the slope of the line defined by the points will not. Same is
true for Figure 1a of the clumps model. The physical scale of density
fluctuations is the same in both the models, so that too can be ruled
out as the source of the difference in the results. The exact form
of the clumps is also not a factor; we have tried a number of profile
shapes, and all yielded very similar results. We are led to conclude that
the difference is due to the shape of the density fluctuations in the clumps 
vs. Gaussian random field models, in particular it is due to the large scale 
coherent, i.e. non-Gaussian nature of the clumps. 

One can arrive at the same conclusion by considering the physical content
of eqs.(11) and (4). According to eq.(4) the second order term introduces 
some extra bending of the light rays towards positive mass inhomegeneities 
and away from negative ones. In a density field dominated by small-scale 
incoherent structures, like our Gaussian random field, the rays are bent 
many times in opposing directions as they travel from the source to the 
observer, thereby averaging to a straight line, a situation well described 
by the standard lensing approach. 

\section{Summary and Conclusions}

We have extended the standard lensing formulation to include the next order
term (second order term in the geodesic equation) in the hope that it will 
make a significant difference in the predicted amplitude of weak lensing 
induced QSO-galaxy correlations, which is currently at least a factor of 5 
smaller than observed. Depending on the type of the mass density fluctuations
populating our model universe, we can obtain up to a $\sim 10\%$ increase in
the amplitude of the correlations, i.e. not enough to reconcile theory with
existing observations. 

Even though the increase is small, qualitative aspects of the updated 
weak lensing formulation bring us closer to the properties of the observed 
correlations. The revised source magnifications correlate with, but are more 
extreme than the magnifications computed using the standard formulation,
i.e. typically $(M_{2nd}-1)/(M_{std}-1)$ is greater than 1, which increases 
the amplitude of predicted lensing-induced correlations. Furthermore, the 
redshift of the optimal lenses for cross-correlations with sources at a 
given redshift is closer to the observer than what the standard formulation 
would predict. This is consistent with observational results. It is important 
to note that the second order term affects only the predictions of the 
lensing theory which deal with weak statistical lensing of point sources. 
As we have shown using eq.(17) and (20) and numerical integrations, image 
elongation, i.e. weak shear lensing, and strong lensing theory are not altered.

Potentially the most important result of the present work is that not all
types of cosmic density fluctuations are sensitive to the presence of the
next higher order term in the lensing equation. Only large-scale, coherent, 
i.e. non-Gaussian structures show an increase in $\omega_{QG}$ with the 
revised formulation. This can be intuitively understood in terms of eq.(11) 
and (4) (see end of previous Section), and quantified with eq.(20), which 
states that the increase in magnification induced by the second order term is 
proportional to a weighted integral of $d\Phi/dz$ over the line of sight.
The integral will tend to be smaller when the phases of the modes comprising
the density and the potential field are uncorrelated, compared to a density
fluctuation model dominated by coherent structures. Note that the increase in 
the predicted cross-correlation, $(\omega_{QG,2nd}/\omega_{QG,std})$ does not 
depend on the amplitude of the mass fluctuations, only on their form.

The work presented here is preliminary, however, if the avenue we have taken
proves to be correct, i.e. if second order, and possibly higher order terms 
in the geodesic equation of motion of photons are important for weak lensing 
induced QSO-galaxy correlations, then these correlations will give us the 
means to study the type of mass structures present on the largest scales, 
in particular to test the Gaussianity of the large scale fluctuations and 
thus shed light on the physical processes in the early Universe that gave 
rise to the present day structures.

\acknowledgements
I am deeply indebted to Prasenjit Saha for numerous invaluable conversations 
on the subject and many helpful suggestions. I am grateful to the anonymous
referee for further elucidating the physical meaning of the second order
term.

\newpage

\begin{figure} 
\plotone{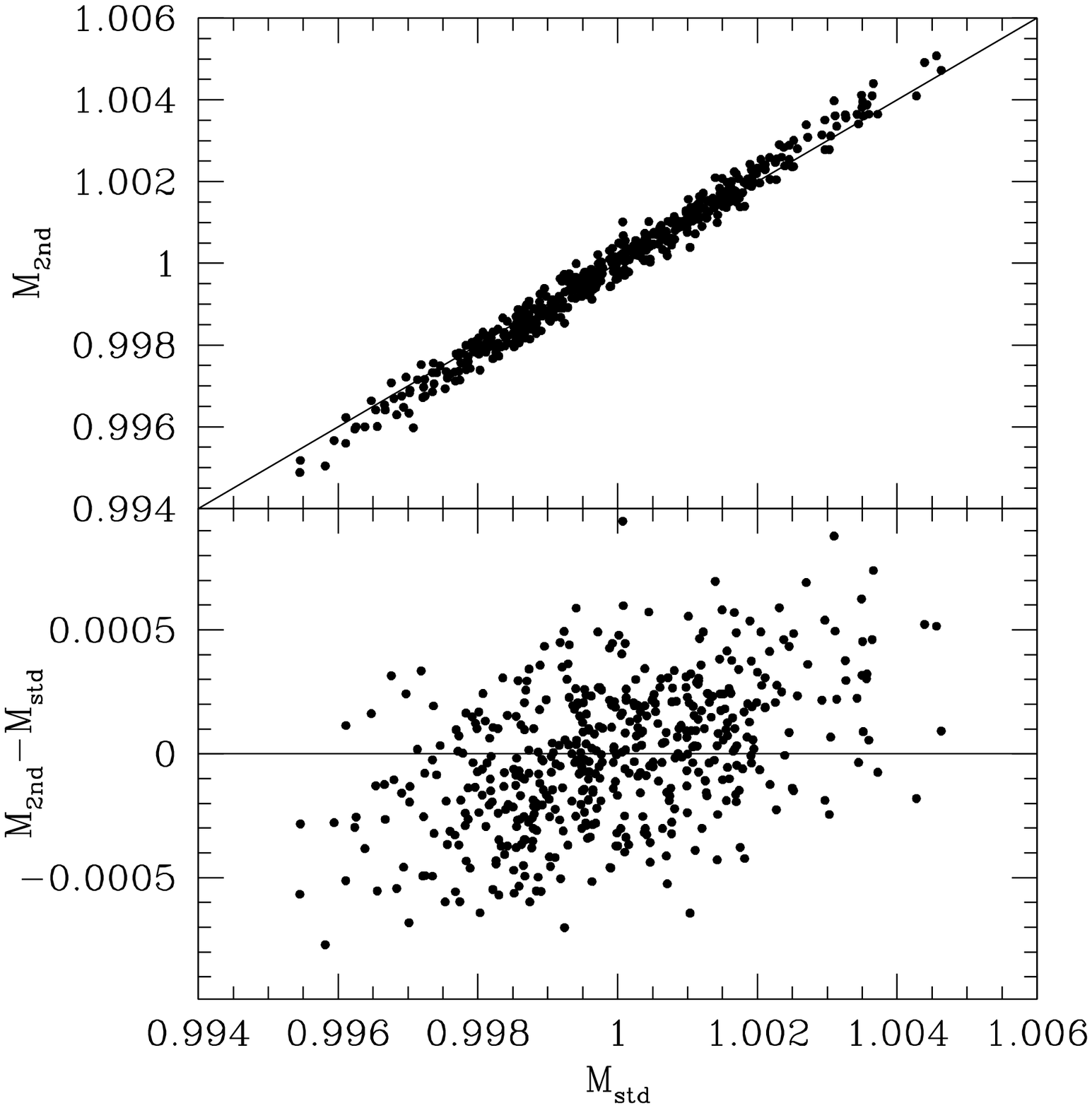}
\figurenum{1} 
\caption{Magnification of sources obtained using the standard lensing
formalism, $M_{std}$, and magnification computed when the second order
term in the geodesic equation is included, $M_{2nd}$. The mass 
fluctuations in this model universe are build up with a collection of 
randomly distributed spherical clumps of positive and negative central 
density. See Section 6 for details. The residuals, $\Delta M$, plotted 
in the bottom panel, are positively correlated with $M_{std}$, i.e.
inclusion of the second order term enhances (de)magnifications.}
\end{figure}  

\begin{figure} 
\plotone{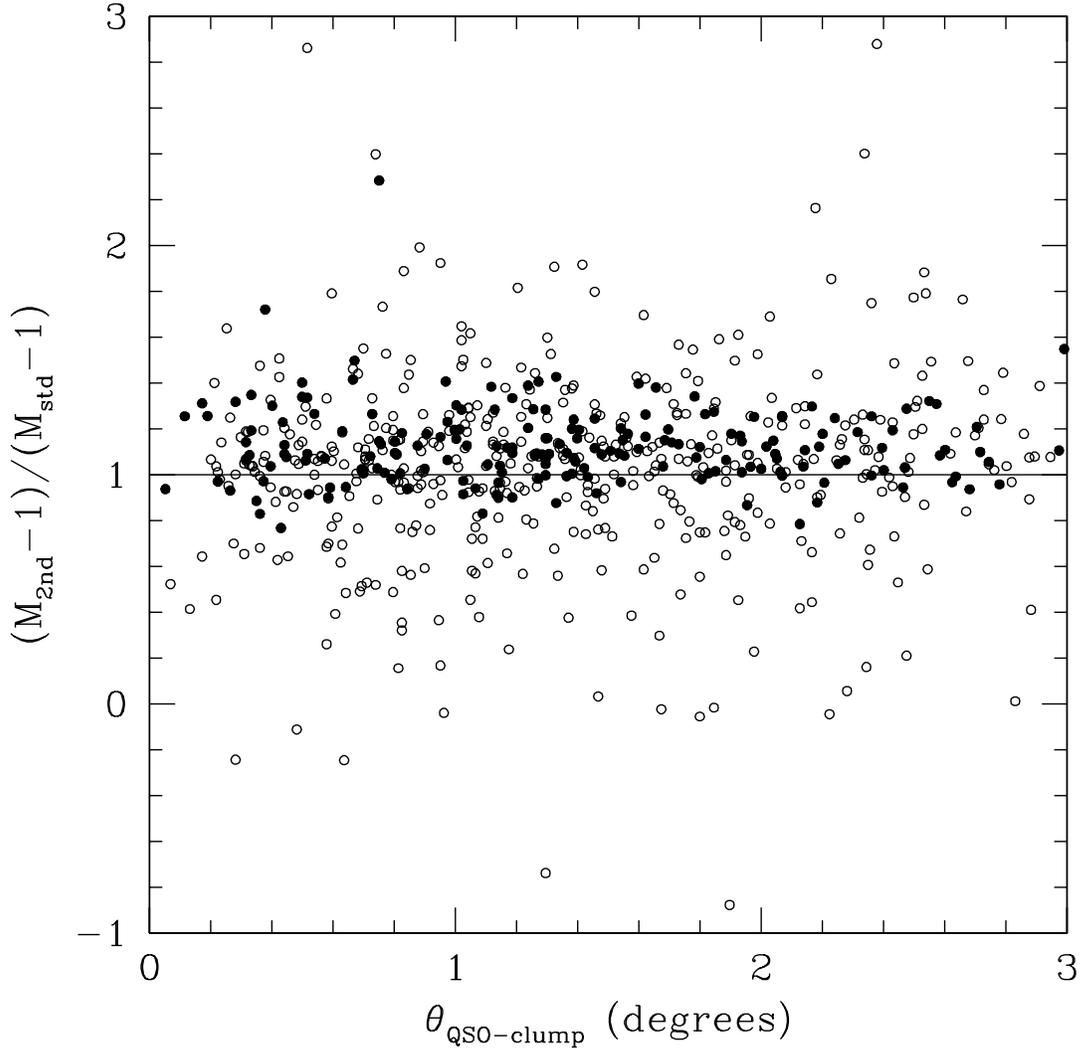}
\figurenum{2} 
\caption{The factor by which cross-correlations are increased is plotted
against the projected distance to the nearest mass clump in the
$z_l=0.1-0.3$ redshift range. All the sources are at $z_s=1.5$, and mass
density fluctuations are represented by a collection of spherically
symmetric mass clumps (see Section 6 for more details.) The median
increase in the amplitude of $\omega_{QG}$ is 1.1. The solid points
are those with $M_{2nd}>1.001$ and have the same median. Correlations
are increased on all angular scales. }
\end{figure}  

\begin{figure} 
\plotone{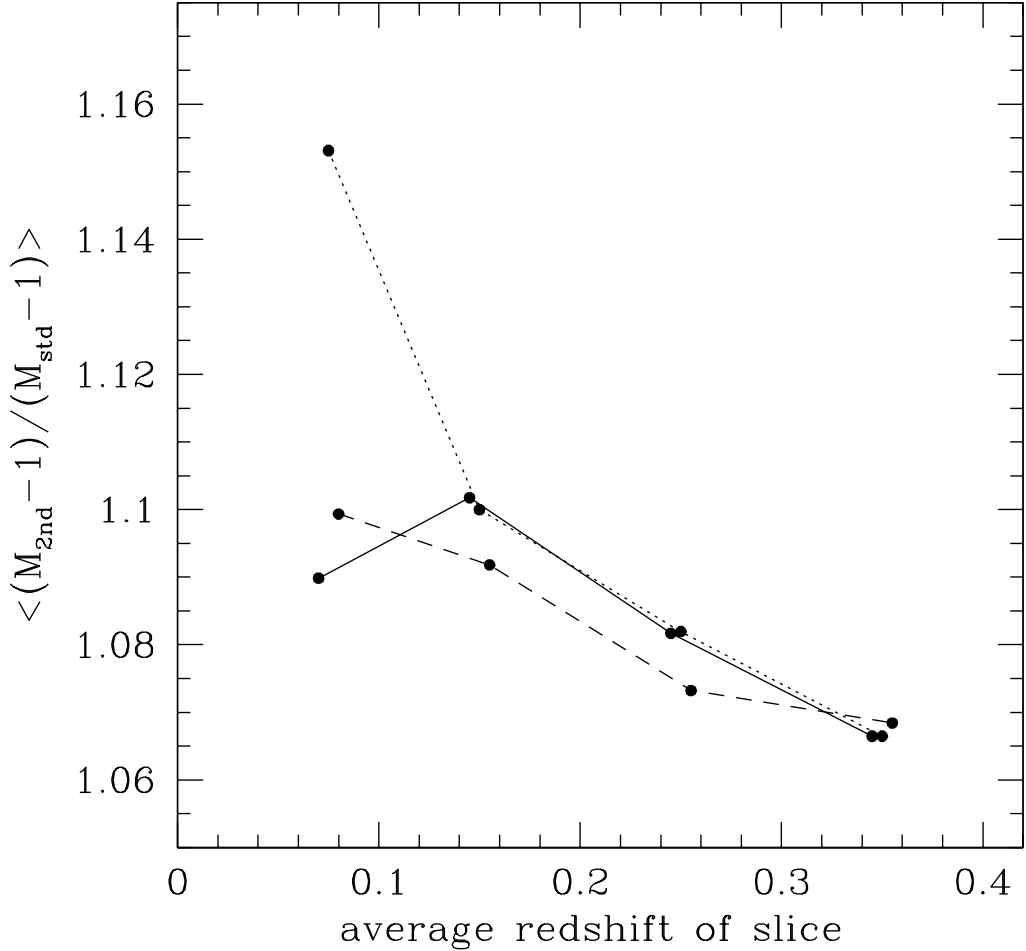}
\figurenum{3} 
\caption{Change in the amplitude of predicted cross-correlation function 
between $z_s=1.5$ sources and mass distribution in four low-redshift slices, 
$z_l=0.05\rightarrow 0.1$, $0.1\rightarrow 0.2$, $0.2\rightarrow 0.3$, and 
$0.3\rightarrow 0.4$. In each of the four redshift slices the median value of
$\langle(M_{2nd}-1)/(M_{std}-1)\rangle$, equal to 
$(\omega_{QG,2nd}/\omega_{QG,std})$, was computed for sources lying within 
$\theta$ of a mass clump (positive central mass density clumps were used 
here). Angle $\theta$ was chosen to be comparable to the angular scale of
observations, $\theta=0.5^\circ$ (solid), $0.75^\circ$ (dotted), and 
$1.0^\circ$ (dashed). (Note that the number of lines of sight contributing to 
the twelve points of the plot varies significantly: the $\theta=0.5^\circ$, 
$z_l=0.05\rightarrow 0.1$ point has 20 lines of sight, while 
$\theta=1.0^\circ$, $0.3\rightarrow 0.4$ point has $\sim 1700$.)
The inclusion of the second order term
in the geodesic equation preferentially increases the amplitude of
correlations for more nearby lenses, thus `moving' the optimal lenses
closer to the observer, compared to the standard optimal lensing distance.}
\end{figure} 

\begin{figure} 
\plotone{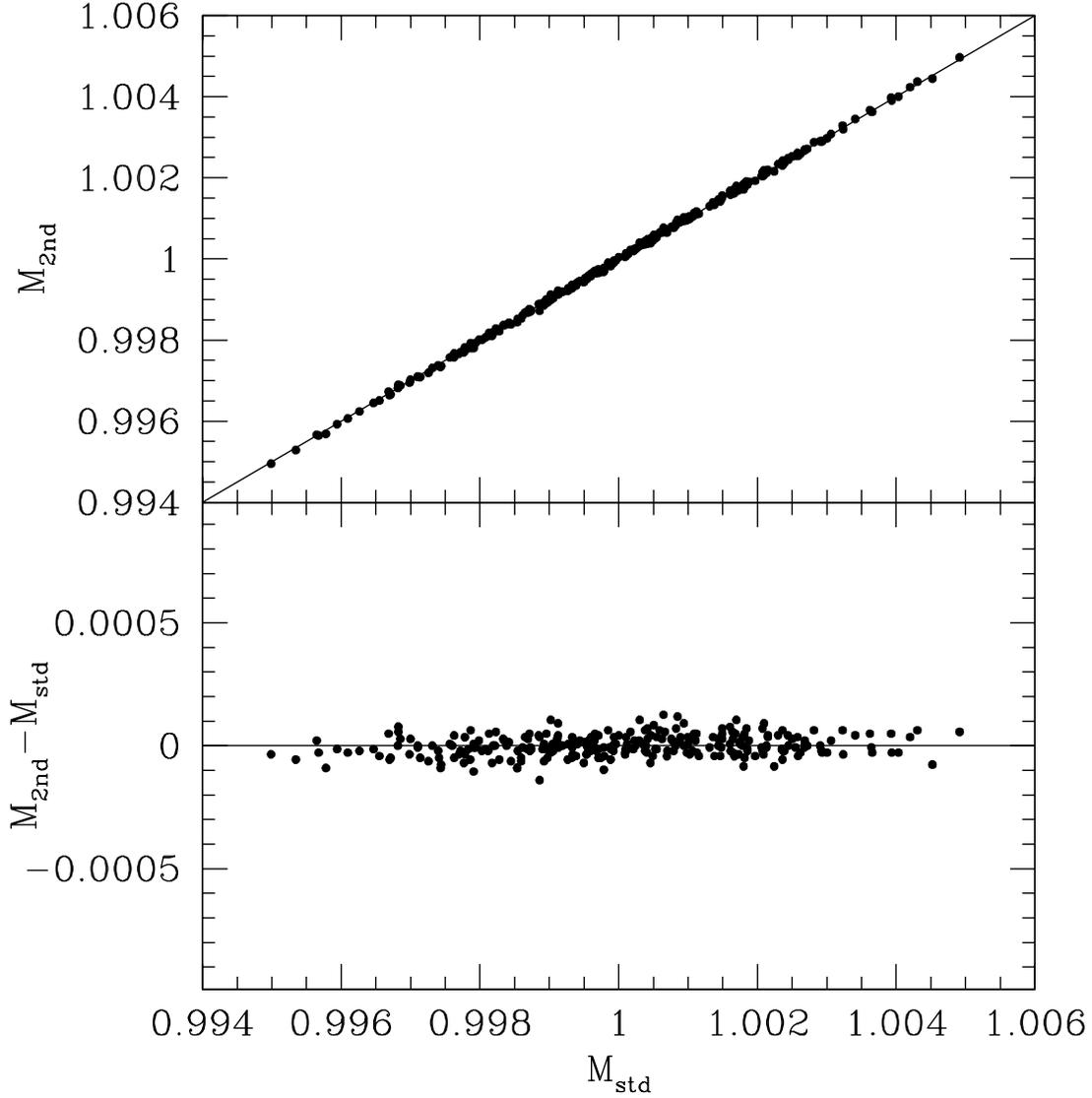}
\figurenum{4} 
\caption{Similar to Figure 1, except the mass fluctuations here are
described by a Gaussian random-phase field, with the power spectrum,
$P(k)\propto k$, and limited to the range of wavenumbers between
$k_{low}=2\pi(2 r_c)^{-1}=0.01h$Mpc$^{-1}$ and 
$k_{high}=2\pi(0.5 r_c)^{-1}=0.04h$Mpc$^{-1}$.
The residuals, $\Delta M$ are much smaller than in Figure 1.}
\end{figure}

\end{document}